\documentclass[prb,aps,showpacs,unsortedaddress,superscriptaddress,amsmath,amssymb,floatfix,twocolumn]{revtex4-1}
\usepackage{epsfig}
\usepackage{graphicx}
\usepackage{amssymb}
\usepackage{amsmath}
\usepackage{amsthm}
\usepackage{wasysym}
\usepackage{hyperref}
\usepackage[usenames]{color}

\begin{document}

\title{Electronic structure of metallic tetra-boride $\textrm{TmB}_{\textrm{4}}$: An LDA+DMFT study}
\author{Nandan Pakhira}
\affiliation{Department of Physics, Indian Institute of Technology, Kharagpur, West Bengal 721302,India}
\affiliation{Department of Physics, Kazi Nazrul University, Asansol, West Bengal 713340,India}
\author{Jyoti Krishna}
\affiliation{Department of Physics, Indian Institute of Technology, Roorkee, Uttarakhand 247667,India}
\author{S. Nandy}
\affiliation{Department of Physics, Indian Institute of Technology, Kharagpur, West Bengal 721302,India}
\author{T. Maitra}
\affiliation{Department of Physics, Indian Institute of Technology, Roorkee, Uttarakhand 247667,India}
\author{A Taraphder}
\affiliation{Department of Physics, Indian Institute of Technology, Kharagpur, West Bengal 721302,India}
\affiliation{Center for theoretical studies, Indian Institute of Technology, Kharagpur, West Bengal 721302,India}
\affiliation{School of Basic Sciences, Indian Institute of Technology Mandi, HP 175001,India}

\begin{abstract}  
Recent experimental observations of magnetization plateau in metallic tetraboride $\textrm{TmB}_{4}$ have created a lot of interest in these class of materials. 
Hysteretic longitudinal resistance and anomalous Hall Effect are other remarkable features in the rare-earth tetraborides which represent experimental realizations 
of Archimedean Shastry-Sutherland (SSL) lattice. Electronic band structures, calculated under GGA and GGA+SO approximations, show that $\textrm{TmB}_{4}$ is a 
narrow band system with considerable correlation in its f-level. Strong correlation effects in this system are studied under single-site dynamical mean field theory 
(DMFT) [LDA+DMFT scheme] using multi-orbital generalization of iterated perturbation theory (MO-IPT). Pseudo-gap behaviour in spectral function and non-Fermi liquid 
behaviour of self-energy shows non-trivial strong correlation effects present in this geometrically frustrated metallic magnets. We also consider the extant, 
heather-to-neglected, strong atomic spin-orbit coupling (SOC) effects. While there is a significant change in the topology of the Fermi surface in the presence of 
SOC, the non-Fermi liquid behavior survives. The system can be modelled by an effective two orbital spinless Falicov-Kimball model together with two free band like 
states.
\end{abstract}
\maketitle
\section{Introduction}
Interplay between geometric frustration and strong correlation effects can lead to exotic quantum phases like quantum spin liquid in frustrated magnets
~\cite{MilaBook} and superconductivity in organic materials~\cite{RossRepProgPhys2011}. While the role of geometric frustration in insulating magnets is well 
studied, its role in metallic systems is not. Rare-earth tetraborides are one such class of materials with chemical formula $\textrm{RB}_{4}$ [$R$ being rare-earth 
atom]. They are frustrated metallic magnets with long range ordered magnetic ground states. It is interesting to mention that the position of the rare-earth atoms 
on two dimensional sheets perpendicular to the tetragonal $c$-axis closely resembles the Archimedian SSL~\cite{SSLmodel}. 

Recent experimental observation~\cite{SaiSwaroopPRB2016} of fractional magnetization plateau at low temperature has drawn significant interest in the insulating 
materials of this series. The magnetization plateaus occurring at a fraction of saturation magnetization with stable plateau at 1/2 and fractional plateaus at 1/7,
1/8, $\cdots$ etc. closely mimic the plateaus observed in the Hall resistivity of a two dimensional degenerate electron gas. The observation of magnetization plateau in $\textrm{TmB}_{4}$ has been attributed to huge spin degeneracy present~\cite{MichimuraPhysicaB,YoshiPRL2008,SiemensmeyerPRL2008,MatasJPhysConf2010} in nearest 
neighbour large spin model on an SSL~\cite{MolinerPRB2009,QinJApplPhys2011,SlavinLTMPhys2011,GrechnevPRB2013}. However, for a metallic system screening effects are 
significant and the localised spins interact via long range RKKY like interactions. Hence the mapping of interacting fermionic model onto an effective spin model on 
SSL with nearest neighbour interaction is highly non-trivial.

The rich phase diagram, hysteretic longitudinal magneto-resistance and anomalous Hall resistivity are other notable experimental observations in $\textrm{TmB}_{4}$
~\cite{SaiSwaroopPRB2016}. First-principles study of the electronic structure of rare-earth tetraborides are few in number. LDA+U calculations of a few of the 
tetraboride systems were performed by Yin et al.~\cite{YinPRB2008}. For $\textrm{TmB}_{4}$, a first-principles calculation showed its band characters recently
~\cite{SP2017}. 

Using first principles (LDA+U approximation) Shin et. al.~\cite{ShinPRB2017} argued that the electronic structure of $\textrm{TmB}_{4}$ admits of a local Kondo-Ising type model. It is important to mention that $\textrm{Tm}$ is nominally in a trivalent state with electronic configuration $4f^{12}$. $4f$ systems are inherently 
narrow band systems and it is well known that electronic correlation effects in such systems are quite strong and cannot be treated within the LDA+U approximation 
where the only effect of $U$ is to shift the local chemical potential. Also it is well known that interplay of strong correlation effects and Hund's coupling in a 
multi-orbital system can lead to unusual effects, like \textit{orbital selective Mott transition} (OSMT)~\cite{AnisimovEPJB2002,deMediciPRB2005,WernerPRL2007,
KogaPRL2004,FerreroPRB2005,LiebschPRL2005,BiermannPRL2005,deMediciPRB2011}. In this paper, using dynamical mean field theory (DMFT)~\cite{GeorgesRMP1996}, we study 
the strong correlation effects on metallic $\textrm{TmB}_{4}$, modelled by realistic bands obtained from density functional theory (DFT). This method is known as 
LDA+DMFT in the literature~\cite{KotliarRMP2006}.                 
\begin{figure*}
\begin{center}
\includegraphics[width=14cm,clip=]{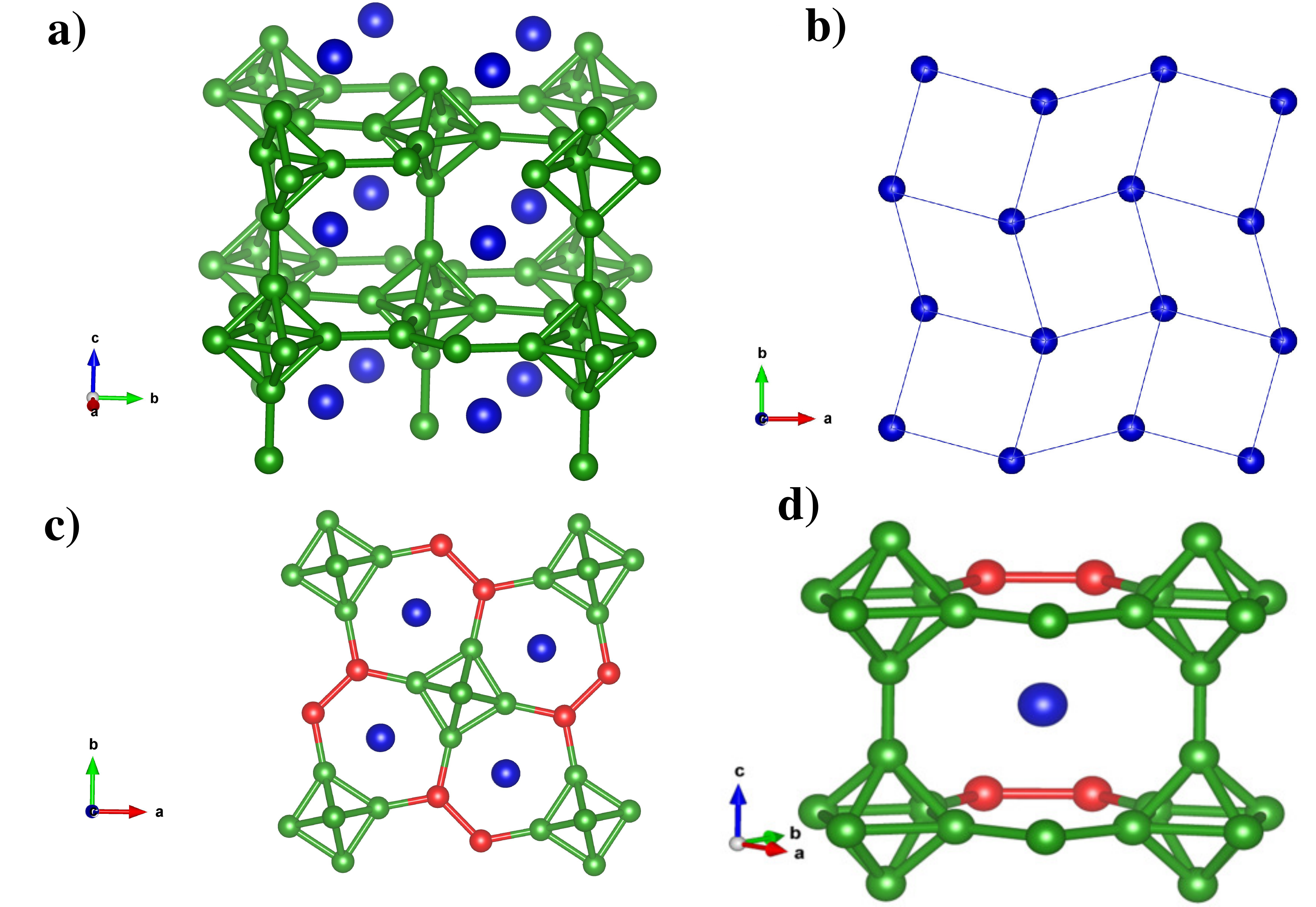}
\caption{(Color online) Tetragonal crystal structure of $\textrm{TmB}_{4}$. Fig.1(a) represents the full structure showing different layers of Tm (blue) and B (green) ions stacked along the c-axis. (b) The Tm ions in the ab plane can be mapped to the Sashtry-Sutherland lattice. (c) The sub lattice of B as viewed along ab plane comprising of 7 atom ring and a square .(d) Showing two different types of B; one forming dimer(shown in red) and other forms a part in B octahedra (shown in green).}
\label{fig: 1}
\end{center}
\end{figure*}

We consider a model Hamiltonian with four spin-degenerate orbitals, developed using density functional theory (DFT) [under GGA], and we study strong electronic 
correlation effects in this narrow band system by applying DMFT with MO-IPT as the impurity solver. Spectral function obtained from self-consistent local Green's 
function consists of a negative energy charge fluctuation peak and broad peaks around the Fermi energy. Spectral features also suggest that for low energy properties we can integrate out one of the orbitals whose integrated spectral weight is predominantly in the positive energy region. Angle resolved photo-emission spectra 
(ARPES) shows significant dispersion along various high symmetry directions except along $X$-$M$ due to flat bands along this direction. The imaginary part of the 
self-energy shows asymmetric behaviour about the Fermi energy. Most importantly, the imaginary part of self-energy for one orbital shows non-Fermi liquid behaviour 
whose origin is possibly linked to pseudo-gap behaviour of corresponding spectral function.
        
In addition, due to large atomic mass of $\textrm{Tm}$, the atomic spin orbit coupling (SOC) effect is quite strong. To the best of our knowledge, the effect of 
strong SOC on the bands of $\textrm{TmB}_{4}$ has not been considered in any previous study. In this paper we include its effect on correlated bands. Due to strong 
SOC effect, the topology of the Fermi surface gets significantly modified. Interestingly even in the presence of SOC we get 3 two-fold degenerate bands. Even more 
interestingly, strong correlation effects and finite Hund's coupling shows degeneracy lifting at low temperature for certain degenerate manifold. The orbitals in 
other manifold still remain degenerate. Finally, non-Fermi liquid characteristics of some bands remain unchanged.

The organization of the rest of the paper is as follows. In Sec. II we discuss crystal structure of the system. In Sec. III we explain computational details for 
band structure. Then we discuss electronic band structure in Sec. IV. In Sec. V we introduce the model involving strong correlation effects and discuss its solution 
under LDA+DMFT scheme. In the same section we describe results on the spectral function, self-energy, ARPES and constant energy surfaces. In Sec. VI we introduce 
the model involving strong correlation and strong spin-orbit coupling effects. We also discuss SOC effects on spectral properties of the model as well as its effect 
on the constant energy surfaces. In Sec.VII we introduce an effective low energy model for the system and briefly discuss its connection with $\textrm{TmB}_{4}$. 
Finally in Sec. VIII we summarize our results and conclude.

\section{Crystal Structure}
$\textrm{TmB}_{4}$ crystallizes in the tetragonal symmetry with space group P4/mbm. It consists of alternate layers of Tm and B ions stacked along c-axis. Fig. 1 
shows the crystal structure of $\textrm{TmB}_{4}$ from different perspectives. Fig.~\ref{fig: 1}(a) displays the full three dimensional tetragonal structure. 
Fig.~\ref{fig: 1}(b) shows only the Tm sublattice in the ab-plane which has the same topology as SSL. Fig.~\ref{fig: 1}(c) shows the top view of the crystal 
structure. There are two distinct types of B ions (1) planar and (2) octahedral present in the structure. Boron atoms form octahedra as well as 7-atom rings as 
shown in Fig.~\ref{fig: 1}(c). In Fig.~\ref{fig: 1} (d) one unit formed by four such B octahedra is shown. B atoms lying in the ab-plane also form dimers (shown in 
red), which are arranged in a regular pattern (see Fig.~\ref{fig: 1}(c)).
\section{Methodology}
We performed first principles density functional theory calculations within GGA and GGA+SO approximations to study the electronic structure of $\textrm{TmB}_{4}$. 
These calculations were carried out using Perdew-Burke-Ernzerhof Generalized Gradient Approximation (PBE-GGA)~\cite{Perdew} exchange-correlation functional within 
the full potential linearized augmented plane wave (FP-LAPW) method as implemented in WIEN2k\cite{Blaha}. Crystal structure was obtained from experiment~\cite{fisk}.  Calculations were performed with 168 $\vec{k}$ points in the irreducible wedge of the Brillouin zone (BZ). The muffin tin radii of Tm and B were taken to be 2.5 
a.u. and 1.52 a.u. respectively and the plane wave expansion in spherical harmonics was considered upto angular momentum quantum number l = 10. Because of the 
presence of localized f-electrons, we have included the spin-orbit coupling (SOC) within the second variational method with scalar relativistic wave functions. We 
have performed the calculations for A-type antiferromagnetic (A-AFM) configuration. Wannier orbital fitting of the DFT bands (which are used as inputs in DMFT 
calculations) are obtained using the interface package WIEN2WANNIER~\cite{wien2wannier} along with the Wannier90~\cite{wannier90} package.

\section{Electronic Structure}
\begin{figure}[htbp]
\centering
\includegraphics[width=8cm,clip=]{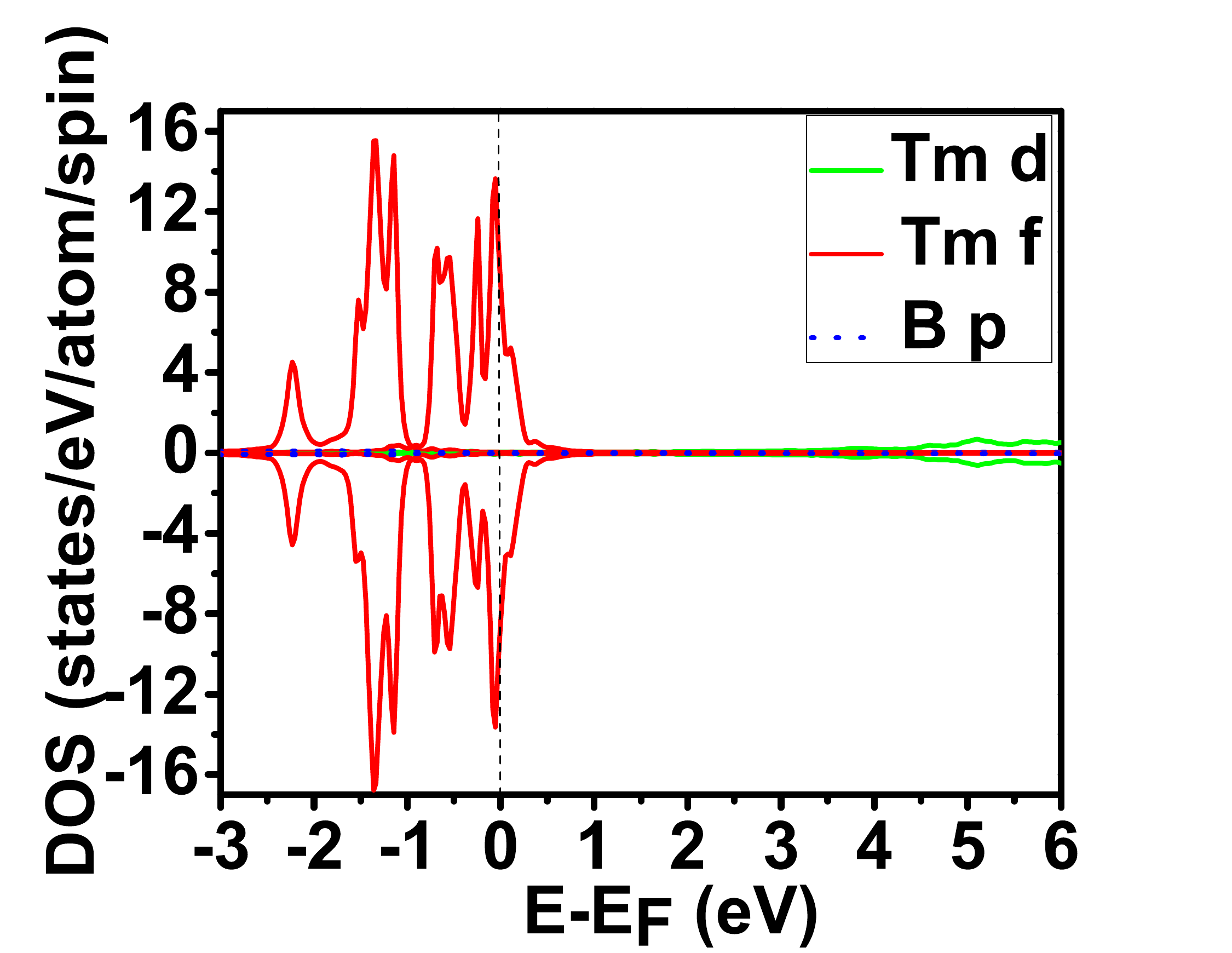}
\caption{(Color online) Spin-polarized partial DOS of Thulium(Tm) $d$,$f$ and Boron(B) $p$ states for both up (UP) and down (DN) spins within GGA approximation. The 
negative DOS is for the DN spin carriers.}
\label{fig: 2}
\end{figure}

\begin{figure*}
\centering
\includegraphics[scale=1.0,clip=]{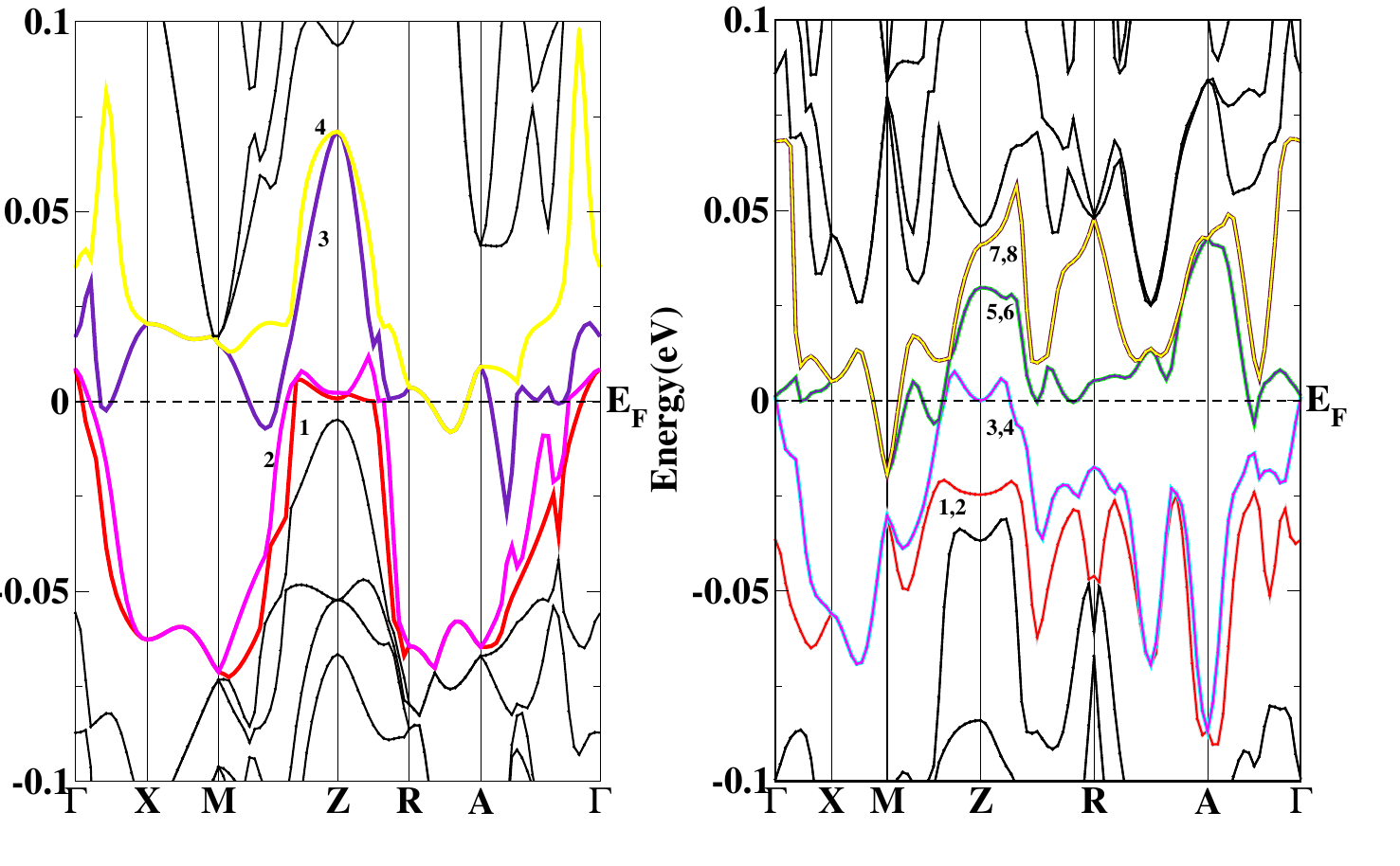}
\caption{(Color online) The band structure for $\textrm{TmB}_{4}$ within GGA (left) and GGA+SO (right). The bands crossing the Fermi energy level ($E_{F}$) are shown 
in colors. Degeneracy at the $M$ point is maintained even in the presence of spin-orbit coupling effect. See text for more details.}
\label{fig: 3}
\end{figure*}

\begin{figure}[htbp]
\centering
\includegraphics[width=8cm,clip=]{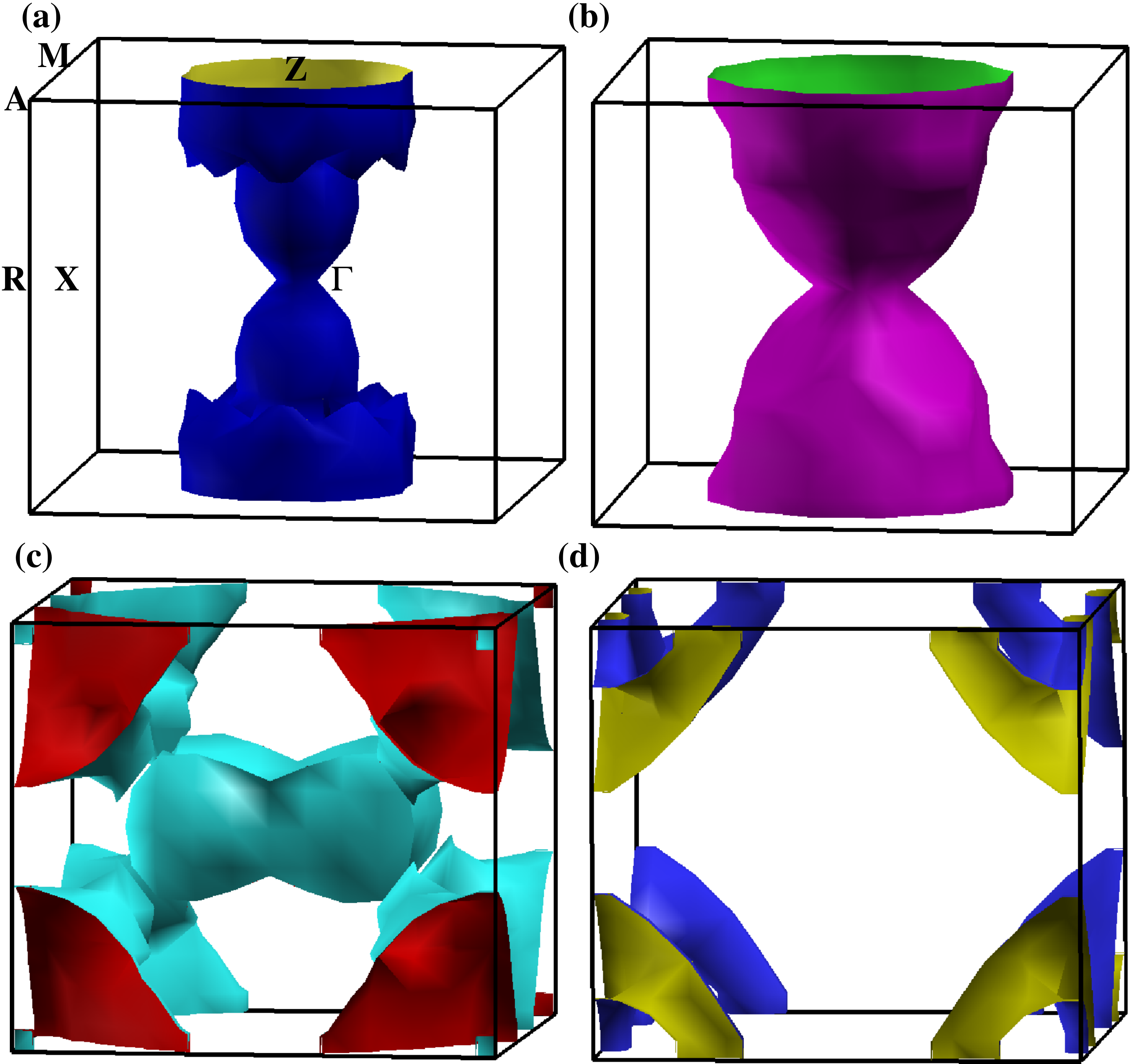}
\caption{(Color online) Panel (a)-(d) : Fermi surface plots for bands crossing the Fermi level within GGA with band index 1,2,3 and 4 [as shown in 
Fig.~\ref{fig: 3}(a)], respectively. The $\Gamma$ point is at the zone centre and various other symmetry points are identified in panel (a).}
\label{fig: 4}
\end{figure}

We discuss below the electronic structure calculated within GGA and GGA+SO approximation for A-type AFM configuration of TmB$_4$ where Tm moments are ferro-
magnetically aligned in the ab-plane and anti-ferromagnetically aligned along c-direction. In Fig.~\ref{fig: 2} we present the spin-polarized partial density of 
states (DOS) of Thulium(Tm) $f,d$ and Boron(B) $p$ states. DOS is clearly metallic in nature with Tm f-states dominantly present at the Fermi level (FL). The 
itinerant carriers from B p and Tm d states also present at the FL and their overlap(hybridization) with Tm f-states make them slightly dispersive which is clearly 
seen in the band structure plots presented in Fig.~\ref{fig: 3}. The band structure calculated within GGA and GGA+SO approximations are plotted along the high 
symmetry path $\Gamma$(0,0,0)-X(0.5,0,0)-M(0.5,-0.5,0)-Z(0,0,-0.5)-R(0.5,0,-0.5)-A(0.5,0.5,-0.5)-$\Gamma$(0,0,0). Within GGA (see Fig 3(a)) four bands cross the FL 
(shown in color) and are mostly of Tm f-character with very small hybridization with B p (mostly along $\Gamma$-X and M-Z-R) and Tm d states. At high symmetry points 
degeneracy is observed among some of these bands. In Fig.~\ref{fig: 4} we present the Fermi surfaces for each of these 4 bands.

With the inclusion of SO interaction, each of these bands crossing FL becomes doubly degenerate and form a set of 8 bands out of which 6 bands (3,4; 5,6; 7,8) now 
cross the FL and remaining two (1,2) are pushed below the FL. We note that at $\Gamma$, along M-Z and at R, the previously seen degeneracy (within GGA) of band 1 
and band 2 has now been lifted. New degeneracies have appeared though. The overall SO splitting has been found to be larger at $\Gamma$, Z and R. The corresponding 
Fermi surfaces are shown in Fig. 5. The Fermi surfaces corresponding to degenerate bands (3,4) consists of disconnected hole pockets along Z-$\Gamma$-Z direction. 
The Fermi surface corresponding to degenerate bands (5,6) consists of hole pockets at $R$ points and electron like pockets and Fermi surfaces in the interior 
region. The Fermi surface corresponding to degenerate bands (7,8) consists of hole pockets at $R$ points. The volume of these hole pockets are much smaller than 
the volume enclosed by the Fermi surfaces corresponding to degenerate bands (3,4) and (5,6). 

\begin{figure}[htbp]
\centering
\begin{tabular}{lll}
\includegraphics[height=12cm,clip=]{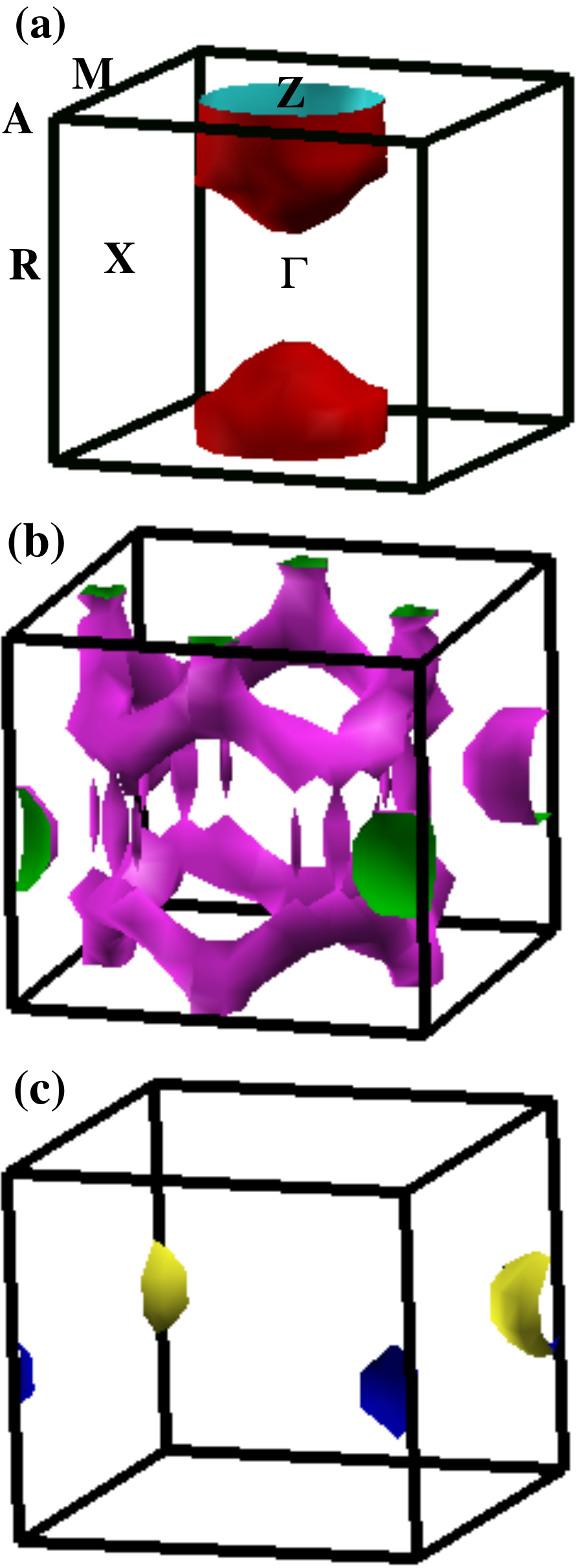}
\end{tabular}
\caption{(Color online) Panel (a) - (c) : Fermi surface plots for bands crossing the Fermi level within GGA+SO with band index 3, 5 and 7 [as shown in 
Fig.~\ref{fig: 3}(b), respectively. The Fermi surfaces for bands with band index 4, 6, and 8 are degenerate with bands with band index 3, 5 and 7 respectively and 
are not shown explicitly.}
\label{fig: 5}
\end{figure}

In all the previously reported electronic structure calculations~\cite{ShinPRB2017,SP2017}, the Coulomb correlation was included within a DFT+U approach whereas here we look at the effect of correlation through DMFT. In a recently reported electronic structure calculation~\cite{SP2017}, the effects of Coulomb correlation and SO interaction were considered within GGA+U and GGA+U+SO approximations.  Due to the application of large U, Tm f-states are pushed about 8-10 eV below Fermi level and only Tm d and B p states straddle the FL. The so called B1 and B3 Boron atoms are seen to have a clear bonding-antibonding splitting of a few eV. The bonding states are nearly filled and anti-bonding states empty. The metallic property is due to the fact that one of the three valence electrons resides in the conduction band. On the other hand Shin et al.~\cite{ShinPRB2017} from their GGA+U calculations find some Tm f-states (strongly hybridized with B p states) are present at the FL giving rise to the metallicity. They further observe that 1.97 holes reside in the Tm f-states giving rise to large hole surfaces of Tm f character and smaller electron surfaces of B p character.
\section{LDA+DMFT}
\begin{figure}
\centering
\includegraphics[scale=0.35,clip=]{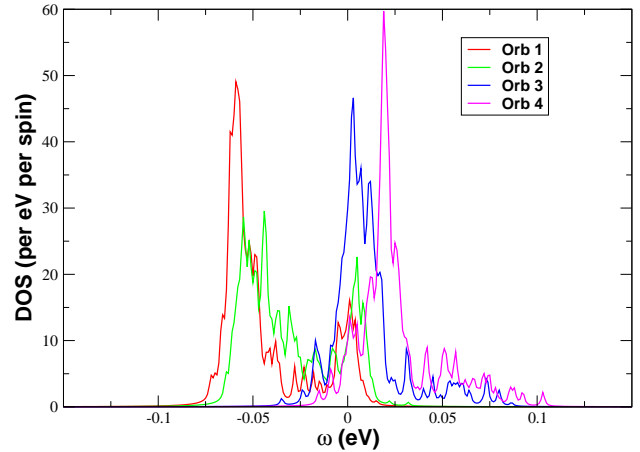}
\caption{(Color online) Partial DOS of 4 Fermi energy crossing bands as shown in Fig.~\ref{fig: 3}. The total band width of each band is less than 0.2 eV.}
\label{Fig:partialDOS}
\end{figure}

The non-interacting Hamiltonian for $\textrm{TmB}_{4}$ can be expressed as
\begin{eqnarray}
H_{0}&=&\sum_{\mathbf{k}\alpha\sigma}(\epsilon_{\mathbf{k}\alpha\sigma}-\mu)c^{\dagger}_{\mathbf{k}\alpha\sigma}c_{\mathbf{k}\alpha\sigma}\nonumber \\
&&+\sum_{\mathbf{k}\alpha\neq\beta}\frac{V_{\alpha\beta}}{2}\left(c^{\dagger}_{\mathbf{k}\alpha\sigma}c_{\mathbf{k}\beta\sigma}+H.c.\right)
\end{eqnarray}
where $\epsilon_{\mathbf{k}\alpha\sigma}$, $\alpha=1,\cdots,4$ are the 4 bands crossing the Fermi-energy and $\sigma=\uparrow,\downarrow$ are the spin indices of 
the electrons in each orbital, $\mu$ is the chemical potential of the system and $V_{\alpha\beta}$ is the residual inter-orbital hybridization which has not been 
included at the level of GGA. For simplicity we take $V_{\alpha\beta}=V$. It is important to mention that non-interacting dispersion 
$\epsilon_{\mathbf{k}\alpha\sigma}$ is obtained by diagonalizing band Hamiltonian, $H(\mathbf{k})$, at each $\mathbf{k}$-point. The band Hamiltonian 
$H(\mathbf{k})$ is obtained by fitting low energy bands around the Fermi-energy to 22 Wannier orbital basis. The non-interacting partial DOS for each of the 4 Fermi 
energy crossing bands are shown in Fig.~\ref{Fig:partialDOS}. From Fig.~\ref{Fig:partialDOS} it is noticeable that total band width of each band is less than 0.2 
eV which makes $\textrm{TmB}_{4}$ a narrow band system. It is well known that in narrow band systems electronic correlation effects are significantly large. We 
incorporate such strong correlation effects under single site dynamical mean field theory (DMFT) approximation. The Hamiltonian describing strong correlation effects under DMFT is given by~\cite{YuFuJPCM2003}
\begin{eqnarray}
H_{U}&=&\sum_{\alpha,\sigma\neq\sigma'}U n_{\alpha\sigma}n_{\alpha\sigma'}+\sum_{\alpha\neq\beta,\sigma\neq\sigma'}U'n_{\alpha\sigma}n_{\beta\sigma'}\nonumber \\
&&+\sum_{\alpha\neq\beta,\sigma}\left(U'-J_{H}\right)n_{\alpha\sigma}n_{\beta\sigma}
\end{eqnarray} 
We choose intra-orbital correlation strength $U=0.1\textrm{ eV}$ which will correspond to $U \sim W/2$, $W$ being the band width of the system. This choice of $U$ 
is appropriate for systems with strong correlation effects. We also choose inter-orbital correlation strength $U'=0.05\textrm{ eV}$, inter-orbital hybridization 
strength $V = 0.02\textrm{ eV}$ and the Hund's coupling energy $J_{H}=0.01\textrm{ eV}$. The chemical potential of the system $\mu$ is fixed from the total particle 
number $n_{\textrm{tot}}=\sum_{\alpha,\sigma} n_{\alpha\sigma}=n^{\textrm{DFT}}$. Where $n^{\textrm{DFT}}$ is the total particle number obtained from DFT 
calculation. For we find $\textrm{TmB}_{4}$ $n^{\textrm{DFT}}=4.0$. We solve the single site impurity problem using multi-orbital generalization of iterated 
perturbation theory (MO-IPT)~\cite{DasariEPJB2016}. MO-IPT has been bench marked against other impurity solver like continuous time quantum monte carlo (CT-QMC) and 
has been shown to reproduce photo-emission spectra in $3d$ transition metal oxide compound $\textrm{SrVO}_{3}$ better than other impurity 
solvers~\cite{DasariEPJB2016}. 
 
\subsection{Spectral function}
\begin{figure}[htbp]
\centering
\includegraphics[scale=0.32,clip=]{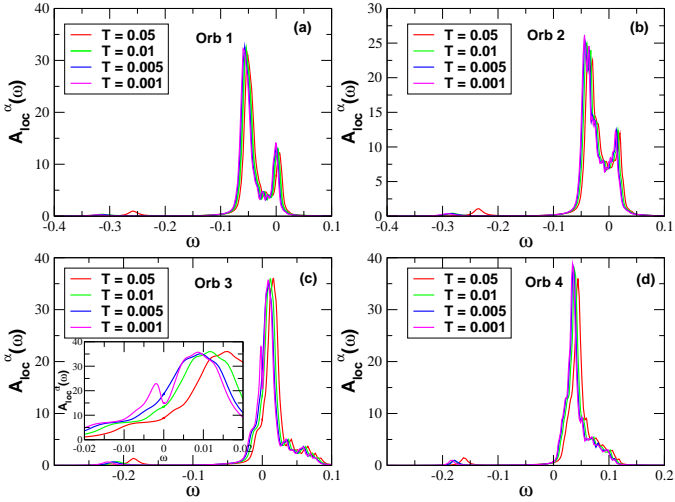}
\caption{(Color online) Local spectral function, $A_{\textrm{loc}}^{\alpha}(\omega)$, for spin degenerate orbital $\alpha$ for various temperatures. The parameters 
chosen are $U=0.1$, $U'=0.05$, $V=0.02$, $J_{H}=0.01$. All energies are measured in units of eV.}
\label{Fig:AlocJH0.01}
\end{figure}

From DMFT self-consistency we obtain local Green's function $G^{\alpha}_{\textrm{loc}}(\omega)$ for each of the orbitals. In Fig.~\ref{Fig:AlocJH0.01} we show the 
corresponding spectral function $A_{\textrm{loc}}^{\alpha}(\omega)\equiv -\textrm{Im}[G^{\alpha}_{\textrm{loc}}(\omega^+)]/\pi$ for each of the 4 degenerate orbitals.
Spectral function of each of the orbitals consists of a low energy charge fluctuation peak and broad peak(s) near Fermi energy. Integrated spectral weight 
under low energy charge fluctuation peak is much smaller than that under broad peaks and strongly temperature dependent - at low temperature spectral weight gets 
transfered to broad peak(s) near Fermi energy. Total bandwidth for broad peak(s) near Fermi energy for each of the orbitals is about $0.15 \textrm{ eV}$. The 
spectral function for orbital 1 and 2 shows two peak structure broadened by correlation effect. Orbital 1 shows a peak at the Fermi energy while orbital 2 shows a 
dip at the Fermi level. With the reduction of temperature the spectral weight gets transfered to lower energy. As a result the peak (orb 1) becomes sharper while 
the dip (orb 2) becomes deeper. This kind of behaviour is expected in multi-band systems with strong correlation effects. The spectral function for orbital 3 and 4 
shows a single peak structure with incoherent spectral features in positive energy. With the reduction of temperature spectral weight gets transfered to lower 
energy as has been observed for the other two orbitals. Most interestingly, at the lowest temperature $T=1\textrm{ meV}$ an additional peak near Fermi energy 
appears for orbital 3 but there is a dip in the spectral function at $\omega=0$ [See inset of Fig.~\ref{Fig:AlocJH0.01}(c)]. The pseudo-gap behaviour of spectral 
function for orbital 3 is possibly a precursor to orbital selective Mott transition (OSMT) state. The behaviour arises mainly due to strong correlation effect in 
narrow band system $\textrm{TmB}_{4}$. Also, finite $J_{H}$ in multi-orbital systems with different band widths can energetically favour OSMT state. For orbital 4 
we see that most of the integrated spectral weight is in the region $\omega > 0$. As a result orbital 4 will not have any significant effect in low temperature dc 
transport and thermodynamic properties of $\textrm{TmB}_{4}$.    
        
\begin{figure}[htbp]
\centering
\includegraphics[scale=0.32,clip=]{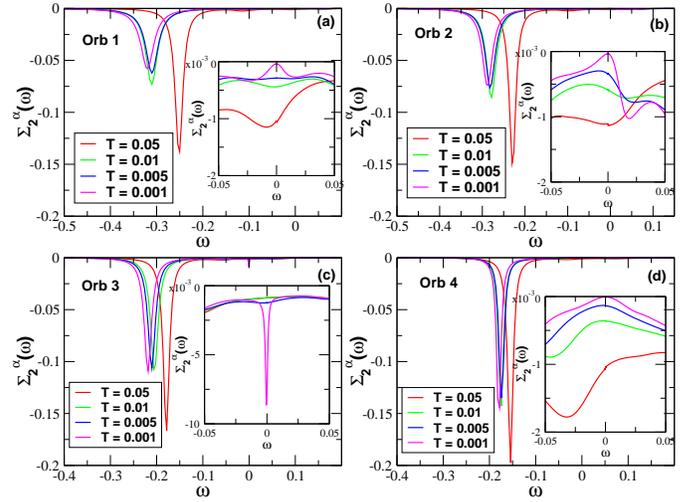}
\caption{(Color online)Imaginary part of impurity self-energy, $\Sigma_{2}^{\alpha}(\omega)$, for spin degenerate orbital $\alpha$ for various temperatures. Inset 
of each plot shows behaviour of $\Sigma_{2}^{\alpha}(\omega)$ in the low frequency region. The parameters chosen are same as in Fig.~\ref{Fig:AlocJH0.01} and all 
the energies are measured in eV. }
\label{Fig:ImSigma}
\end{figure}
\subsection{Self Energy}
From the DMFT self consistency we also get self energy $\Sigma_{\alpha}(\omega)$ for each of the orbitals. In Fig.~\ref{Fig:ImSigma} we show the imaginary part of 
self consistent self energy, $\Sigma_{2}^{\alpha}(\omega)\equiv \textrm{Im}[\Sigma^{\alpha}(\omega)]$, for different orbitals labeled by orbital index $\alpha$ for 
various temperatures. $\Sigma_{2}^{\alpha}(\omega)$ for all the 4 orbitals shows peak in the negative energy region corresponding to the charge fluctuation peak 
observed in the local spectral function $A_{\textrm{loc}}^{\alpha}(\omega)$. Inset of each plot shows low frequency behaviour of $\Sigma_{2}^{\alpha}(\omega)$. 
$\Sigma_{2}^{\alpha}(\omega)$ for orbital 4 shows Fermi-liquid behaviour : $\Sigma_{2}^{\alpha}(\omega)\propto -[\omega^{2}+(\pi k_{B}T)^{2}]$ at the lowest 
temperature $T=1\textrm{ meV}$. For orbital 1 and 2 $\Sigma_{2}^{\alpha}(\omega)$ is asymmetric about $\omega=0$ and there is possible finite intercept at the Fermi 
energy. Most importantly, the low energy behaviour of $\Sigma_{2}^{\alpha}(\omega)$ for orbital 3 shows behaviour contrary to what one would expect for a coherent 
Fermi liquid state. Inset of Fig.~\ref{Fig:ImSigma} (c) clearly shows that $\Sigma_{2}^{\alpha}(\omega)$ for orbital 3 is finite at $\omega=0$ instead of vanishing 
for a coherent Fermi liquid state. Also the slope of $\Sigma_{2}^{\alpha}(\omega)$ is opposite to what one would expect for a Fermi liquid state. This kind of 
behaviour is possibly linked to pseudo-gap behaviour of spectral function for orbital 3 as has been discussed in the previous section. The non-Fermi liquid 
behaviour of $\Sigma_{2}^{\alpha}(\omega)$ is possibly a precursor to OSMT in narrow band multi-orbital system such as $\textrm{TmB}_{4}$.      
\begin{figure}[htbp]
\centering
\includegraphics[scale=0.35,clip=]{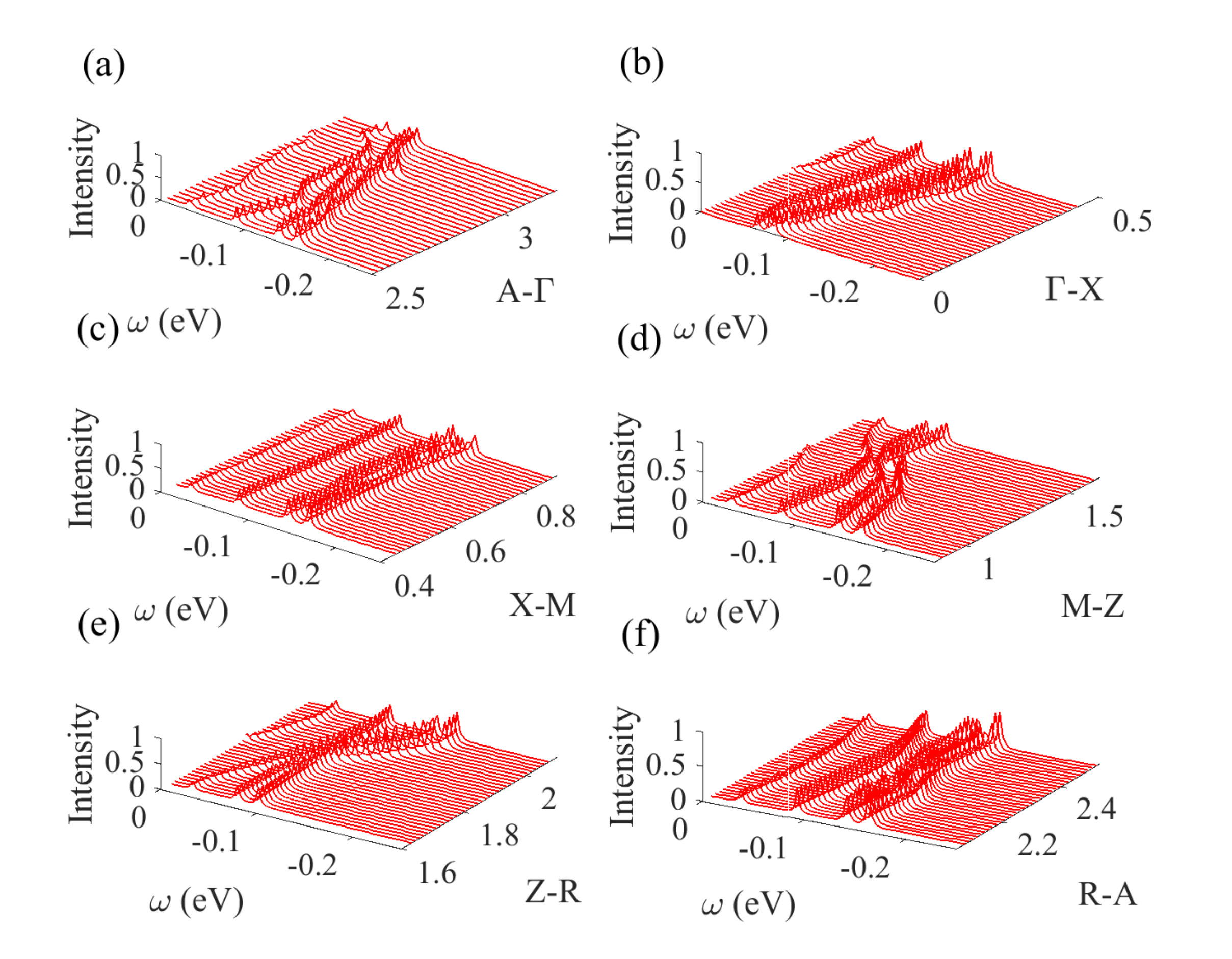}
\caption{(Color online) ARPES intensity [in arbitary units] spectra along the high symmetry directions : (a) A - $\Gamma$ (b) $\Gamma$ - X (c) X - M (d) M - Z 
(e)Z - R (f) R - A at high temperature, $T=0.05\textrm{ eV}$. The parameters chosen are same as in Fig.~\ref{Fig:AlocJH0.01}.}
\label{Fig:ARPES_T0.05}
\end{figure}
\begin{figure}[htbp]
\centering
\includegraphics[scale=0.35,clip=]{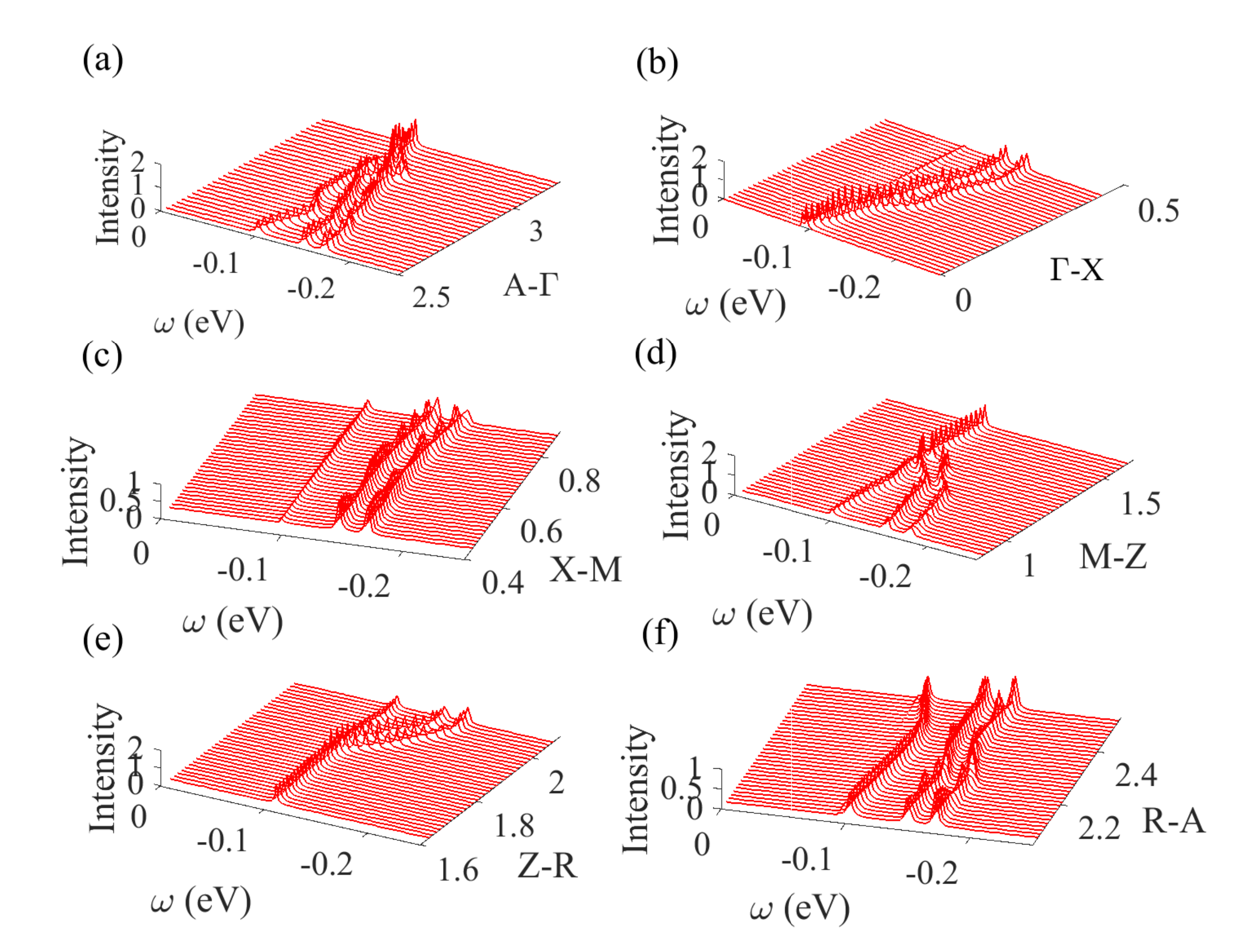}
\caption{(Color online) ARPES intensity [in arbitary units] spectra along the high symmetry directions : (a) A - $\Gamma$ (b) $\Gamma$ - X (c) X - M (d) M - Z 
(e)Z - R (f) R to A at low temperature, $T=0.001\textrm{ eV}$. The parameters chosen are same as in Fig.~\ref{Fig:AlocJH0.01}.}
\label{Fig:ARPES_T0.001}
\end{figure}
\subsection{Angle resolved photo-emission spectra}
In order to understand spectral weight transfer due to strong correlation effects in $\textrm{TmB}_{4}$ we study angle resolved photo-emission spectra (ARPES) along 
high symmetry directions in the first Brillouin zone. ARPES intensity spectrum, $I(\mathbf{k},\omega)$, can be calculated as 
\begin{eqnarray}
I(\mathbf{k},\omega) \propto -\sum_{\alpha}\frac{1}{\pi}\textrm{Im}\left[\frac{n_{F}(\omega)}{\omega^{+}-\epsilon_{\mathbf{k}\alpha}-\Sigma_{\alpha}(\omega)}\right]
\end{eqnarray} 
where $n_{F}(\omega)$ is the Fermi function, $\Sigma_{\alpha}(\omega)$ is the self-consistent self energy from DMFT and $\epsilon_{\mathbf{k}}\alpha$ is the band 
dispersion for orbital $\alpha$ along high symmetry direction. In Fig.~\ref{Fig:ARPES_T0.05} we show ARPES spectra, calculated at $T=0.05\textrm{ eV}$, along high 
symmetry directions in the first Brillouin zone. From the ARPES spectra it is noticeable that spectral peaks are much less dispersive in certain symmetry directions 
like X to M (Fig.~\ref{Fig:ARPES_T0.05}(c)) and between R and A (Fig.~\ref{Fig:ARPES_T0.05}(f)). This features can be understood from the GGA band structure along 
those symmetry directions. From X to M and R to A we have two fold degenerate bands well separated in energy (see Fig. 3). As result spectral weight transfer between 
different bands is largely suppressed giving rise to less dispersive spectral features along respective symmetry directions. On the other hand between $\Gamma$ to 
$X$ and between $M$ to $R$ bands are non-degenerate and appearance of different spectral peaks is due to charge transfer between various non-degenerate bands which 
are closely spaced in energy.

In Fig.~\ref{Fig:ARPES_T0.001} we show ARPES spectra at the lowest temperature $T=1\textrm{ meV}$ calculated along various symmetry directions as described in the 
previous paragraph. Most important noticeable feature of the low temperature ARPES spectra is sharpening of spectral peaks along all the symmetry directions of the 
first Brillouin zone. This is due to reduced self energy effects at lower temperatures. Also, as we compare with ARPES spectra in Fig.~\ref{Fig:ARPES_T0.05} 
disappearance of some of the spectral peaks are clearly visible. The origin of these spectral features at high temperatures are due to charge transfer between 
thermally excited states in different bands. The peak at $\omega\sim -0.1\textrm{ eV}$ can be identified as the spectral peak at $\omega\sim 0.0$ shifted by the 
chemical potential, $\mu$. This peak is found to be less dispersive over the entire Brillouin zone except in the $\Gamma$-$X$ direction. This corresponds to spectral 
weight transfer from $\Gamma$ point to $X$-point. The two closely placed peaks at $\omega\sim -0.15\textrm{ eV}$ and $\omega\sim -0.17\textrm{ eV}$ are due to 
orbital 1 and orbital 2. The $\omega\sim 0.05 \textrm{ eV}$ peak due to orbital 4 gets suppressed by the Fermi factor at low temperature but appears as a 
$\omega\sim -0.05 \textrm{ eV}$ bump at high temperature $T=0.05\textrm{ eV}$.   
\subsection{Constant energy surface}
It is also important to understand the strong correlation effects on the energies of the many-body quantum states. It can be shown that under single site DMFT 
approximation the energies of the many-body states, $\varepsilon_{\alpha}$, in orbital $\alpha$ is related to the energies of the corresponding non-interacting 
states, $\varepsilon_{0\alpha}$, through 
\begin{eqnarray}
\varepsilon_{\alpha}=\varepsilon_{0\alpha}-\mu+\textrm{Re}\left[\Sigma_{\alpha}(\varepsilon_{\alpha})\right],
\label{eqn:E_alpha}
\end{eqnarray}
where $\Sigma_{\alpha}(\varepsilon_{\alpha})$ is the selfconsistent self-energy corresponding to orbital $\alpha$.

We know that in the Fermi liquid state the system can be characterized by weakly interacting quasi-particle states at the Fermi surface. In order to understand 
strong correlation effect on these quasi-particle states at the Fermi level we calculate correlated Fermi surface and compare against the noninteracting Fermi 
surface. In particular we can calculate interacting Fermi surface corresponding to orbital $\alpha$ from the solution of
\begin{eqnarray}
\varepsilon_{\alpha}+\mu-\textrm{Re}\left[\Sigma_{\alpha}(\varepsilon_{\alpha})\right]=0,
\label{eqn:Ezero}
\end{eqnarray}
as the constant energy surface $E_{\mathbf{k}\alpha}=\varepsilon_{\alpha}$. It is important to mention that in a non-Fermi liquid state we can still solve 
Eq.~\ref{eqn:E_alpha} and define a constant energy surface. We are interested in particular constant zero energy surface as the solution of Eq.~\ref{eqn:Ezero}. In 
the weakly correlated regime the zero energy surface evolves into interacting Fermi surface.

\begin{figure}[htbp]
\centering
\includegraphics[width=8cm,clip=]{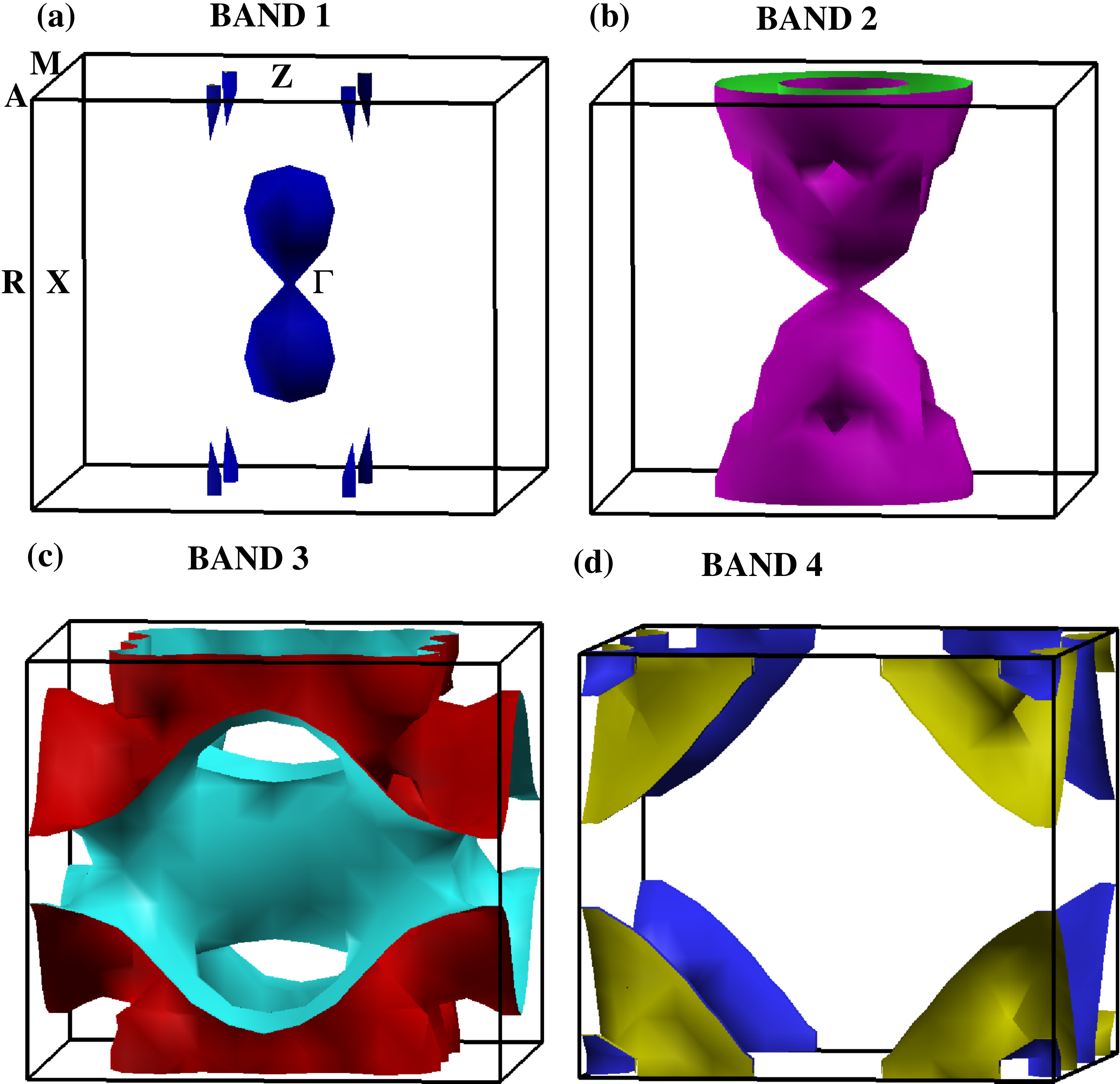}
\caption{(Color online) Interacting Fermi surface for various orbitals (bands) in the first Brillouin zone. Here $T=1\textrm{ meV}$ and all 
other parameters are same as in Fig.~\ref{Fig:AlocJH0.01}.}
\label{Fig:FS_int}
\end{figure}
In Fig.~\ref{Fig:FS_int} we show the constant zero energy surface for various orbitals (bands) in the first Brillouin zone. Appearance of hole pockets and shrinkage 
of Fermi surface volume are the most important noticeable features of the interacting Fermi surface for orbital 1 (band 1). The interacting Fermi surface for orbital 
2 (band 2) gets distorted but its topology remains largely preserved. It is interesting to mention that strong correlation effects have negligible effect on orbital 
4 (band 4). However due to strong correlations and inter orbital hybridization effects the topology of the zero energy surface for orbital 3 (band 3) gets 
significantly affected [see Fig.~\ref{Fig:FS_int}]      
\section{Spin orbit Coupling Effect}
Due to quite large atomic mass of $\textrm{Tm}$ ions spin-orbit coupling (SOC) effect in $4f$ systems like $\textrm{TmB}_{4}$ is quite important. It is important to 
mention that in previous studies of rare-earth tetra-borides SOC effects have been neglected altogether. SOC effect on GGA band structure is shown in the right 
panel of Fig.3. It is important to mention that in the presence of SOC effect spin and orbital degrees of freedoms gets coupled and hence electron spin is no longer 
a good quantum number. We now consider a 8 orbital non-interacting Hamiltonian
\begin{eqnarray}
H_{0}^{SO}&=&\sum_{\mathbf{k}\alpha} \left(\epsilon_{\mathbf{k}\alpha}^{SO}-\mu^{SO}\right)d_{\mathbf{k}\alpha}^{\dagger}d_{\mathbf{k}\alpha}\nonumber \\
&&+\sum_{\mathbf{k},\alpha\neq\beta}V^{SO}_{\alpha\beta}\left(d_{\mathbf{k}\alpha}^{\dagger}d_{\mathbf{k}\beta}+H.c.\right) 
\end{eqnarray}
where $d_{\mathbf{k}\alpha}^{\dagger}$ creates an electron in orbital $\alpha$ at $\mathbf{k}$ in the first Brillouin zone and $\epsilon_{\mathbf{k}\alpha}^{SO}$ is 
the non-interacting band dispersion in the presence of SOC. It is important to mention that $\epsilon_{\mathbf{k}\alpha}^{SO}$ for $\alpha=1,2$ corresponds to two 
fold degenerate orbitals labeled (1,2) below the Fermi level as shown in Fig. 3. We consider these two bands because of their degeneracy with other bands at the $M$ 
point. As a result any residual inter-orbital hybridization can cause spectral weight transfer to other orbitals. $\epsilon_{\mathbf{k}\alpha}^{SO}$ is obtained by 
diagonalizing a Hamiltonian $H^{SO}_{\mathbf{k}}$ at each $\mathbf{k}$-point. $H^{SO}_{\mathbf{k}}$ is obtained by fitting low energy bands around Fermi level to 44 
Wannier orbital basis. $\mu^{SO}$ is the new chemical potential of the system in the presence of SOC. We choose $V^{SO}_{\alpha\beta}=V^{SO}$ if $\beta > \alpha+1$ 
and $V^{SO}_{\alpha\beta}=0$ otherwise. This choice of $V^{SO}_\alpha\beta$ ensures hybridization between widely separated orbitals only.

We include electron-electron correlation effect using a eight orbital Hamiltonian given by~\cite{NandyPRB2016}
\begin{eqnarray}
H_{U}^{SO}=\sum_{\alpha,\alpha'}U_{\alpha\alpha'}n_{\alpha}n_{\alpha'}+\sum_{\alpha\beta} U_{\alpha\beta}n_{\alpha}n_{\beta}
\end{eqnarray}
where $U_{\alpha\alpha'}$ is the interaction energy between electrons in two degenerate orbitals with orbital indices $\{\alpha=1,3,5,7 ; \alpha'=\alpha+1\}$ while 
$U_{\alpha\beta}$ is the interaction energy between electrons in widely separated orbitals. It is well known that interaction energy between electrons in degenerate 
orbitals is stronger than that between electrons in widely separated orbitals. For notational convenience we choose $U_{\alpha\alpha'}=U$ for the rest of the paper. 
We choose $U_{\alpha\beta}$ non-zero only if $\beta > \alpha+1$. Further $U_{\alpha\beta}=U'$ if the pair of indices $\alpha,\beta$ are different kind i.e. $\alpha$ 
being odd and $\beta$ even and vice versa and $U_{\alpha\beta}=U'-J_{H}^{SO}$ if the pair of indices $\alpha,\beta$ are of same kind i.e. both being odd or even. Inclusion of $J_{H}^{SO}$ closely mimics the role of $J_{H}$ between spin degenerate orbitals in atomic physics. Inclusion of $J_{H}^{SO}$ reduces some of the 
correlation effects and hence promotes gain in the kinetic energy and metallicity. We study the effective Hamiltonian $H_{\textrm{eff}}^{SO}=H_{0}^{SO}+H_{U}^{SO}$ 
under single site DMFT approximation and MO-IPT is the impurity solver. We choose $U=0.1\textrm{ eV}$, $U'=0.05\textrm{ eV}$, $V^{SO}=0.02\textrm{ eV}$ and 
$J_{H}^{SO}=0.01\textrm{ eV}$.
\subsection{Spectral function}  
\begin{figure}[htbp]
\centering
\includegraphics[scale=0.32,clip=]{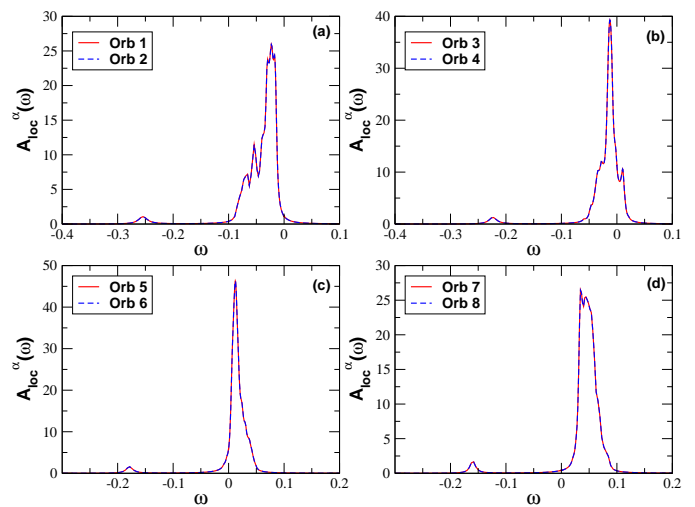}
\caption{(Color online)Spin-orbit (SO) coupling effect on local spectral function, $A_{\textrm{loc}}(\omega)$, at high temperature $T=0.05$. Degeneracy lifting 
effect is compensated by the thermal broadening effect. The parameters chosen are $U=0.1\textrm{ eV}$, $U'=0.05\textrm{ eV}$, $V^{SO}=0.02\textrm{ eV}$ and 
$J_{H}^{SO}=0.01\textrm{ eV}$[see text for details] and all energies are measured in eV.} 
\label{Fig:AlocSOT0.05}
\end{figure}
\begin{figure}[htbp]
\centering
\includegraphics[scale=0.32,clip=]{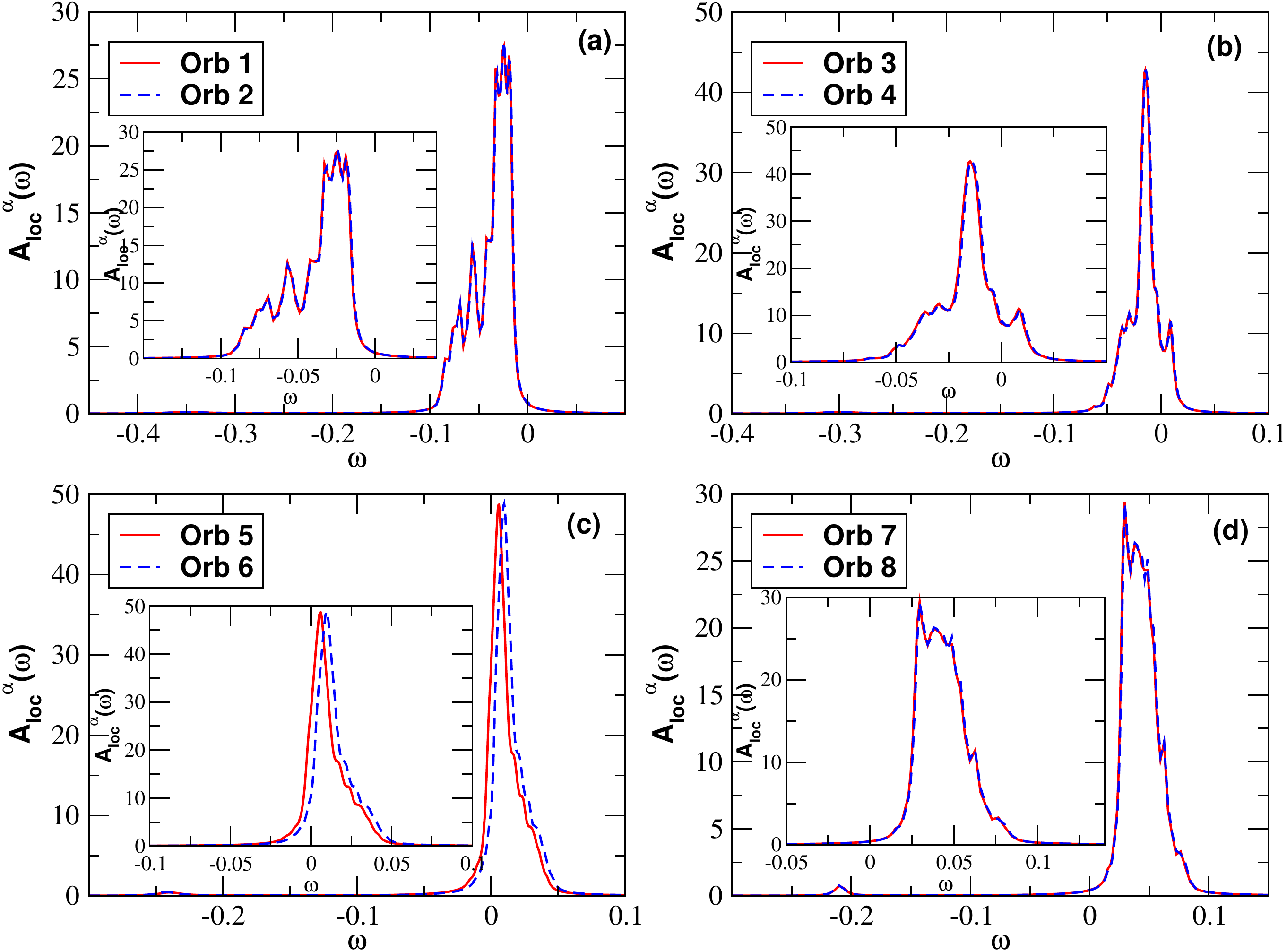}
\caption{(Color online) SO coupling effect on local spectral function, $A_{\textrm{loc}}(\omega)$, at low temperature $T=0.001$. Inset of each plot shows blown up 
central peak. Degeneracy lifting effect is quite noticeable especially for orbitals 5 and 6. The parameters are same as in Fig.~\ref{Fig:AlocJH0.01} and all 
energies are measured in units of eV.}
\label{Fig:AlocSOT0.001}
\end{figure} 
In Fig.~\ref{Fig:AlocSOT0.05} we show SOC effect on the spectral function at high temperature $T=0.05\textrm{ eV}$. In each panel we show spectral function for 
degenerate orbital pairs. The most important feature of all the plots is that the degenerate orbital pairs remain degenerate. This is possibly due to thermal 
broadening effects compensating any possible degeneracy lifting effects. Spectral function for each of the degenerate pairs of orbitals consists of peaks around 
Fermi energy and a charge fluctuation peak around $\omega\sim -0.2\textrm{ eV}$. Orbital 1 and 2 are completely filled and will have no significant contribution 
to thermodynamic and dc transport properties. For orbital 3 and 4 we see two sharp peaks around Fermi energy and incoherent features extending upto 
$\omega\sim -0.075\textrm{ eV}$. The negative energy peak is quite sharp while the positive energy peak is less prominent and more of a sub peak also there is a 
dip in the density of states at the Fermi level. For orbital 5 and 6 there is a single peak around Fermi energy and the band width for this peak is smaller 
than all other orbitals. Spectral function for orbital 7 and 8 consists of a broad peak in the positive energy with a possible splitting due to SOC effect. 

In Fig.~\ref{Fig:AlocSOT0.001} we show SOC effect on the spectral function at low temperature $T=0.001\textrm{ eV}$. As in the case without SOC effect the 
negative energy charge fluctuation peak gets strongly suppressed and the spectral weight under this peak is transfered to the region around Fermi energy. For all 
other orbitals except orbital 5 and 6 there is no significant change in the spectral features except the peaks get sharper and there is appearance of additional 
kinks on the spectral function. The most significant effect of SOC effect is observed in spectral function for orbital 5 and 6. There is clear signature of 
degeneracy lifting between orbital 5 and 6. It is important to mention that interplay between SOC effect and Hund's coupling leads to this degeneracy lifting 
effect. The degeneracy lifting between orbital 5 and 6 can give rise to magnetic instability at low temperature.          
\subsection{Selfenergy}  
\begin{figure}[htbp]
\centering
\includegraphics[scale=0.32,clip=]{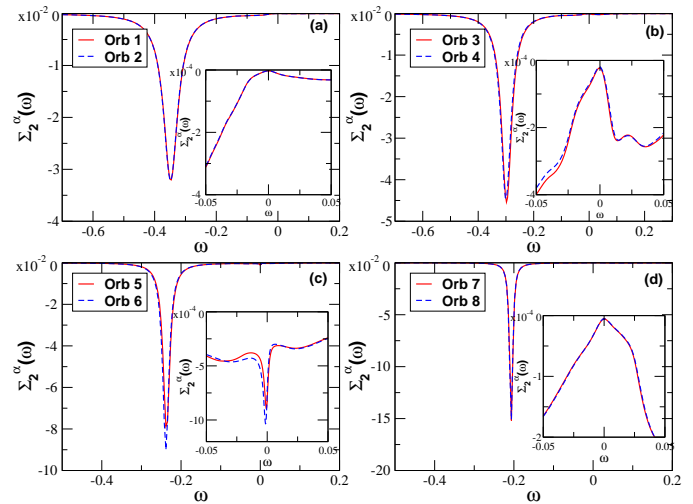}
\caption{(Color online) SO coupling effect on imaginary part of self-energy, $\Sigma_{2}(\omega)$, at low temperature $T=0.001$. Inset of each plot shows low 
frequency behaviour of $\Sigma_{2}(\omega)$. Degeneracy lifting effect is most prominent for orbital 5 and 6 (see text for details). The parameters are same as in 
Fig.~\ref{Fig:AlocSOT0.05} and all energies are measured in units of eV.}
\label{Fig:ImSigma_SO}
\end{figure}
In the last section we showed that at low temperature SOC effect can lift degeneracy between certain orbital pairs (orbital 5 and 6). Therefore it is important 
and interesting to understand SOC effect on the self energy of degenerate orbital pairs at low temperature. In Fig.~\ref{Fig:ImSigma_SO} we show SOC effect on the 
imaginary part of self energy, $\Sigma_{2}^{\alpha}(\omega)$, of degenerate orbital pairs at $T=1\textrm{ meV}$. Imaginary part of self energy for all 
degenerate orbital pair is dominated by a peak in the negative frequency region. This peak corresponds to charge fluctuation peak observed at around 
$\omega\sim -0.2\textrm{ eV}$. The inset of each panel shows detail behaviour of $\Sigma_{2}^{\alpha}(\omega)$ in the low frequency region. Except for orbital pair 
5 and 6 no other degenerate orbital pairs show any significant degeneracy lifting effect due to SOC. $\Sigma_{2}^{\alpha}(\omega)$ for all orbital pairs are 
highly asymmetrical about $\omega=0$ with possible finite intercept at the Fermi level. $\Sigma_{2}^{\alpha}(\omega)$ for orbital 5 and 6 shows degeneracy 
lifting effect as expected but the behaviour of $\Sigma_{2}^{\alpha}(\omega)$ at $\omega=0$ is contrary to Fermi liquid characteristics - 
$\Sigma_{2}^{\alpha}(\omega)$ is finite at $\omega=0$ instead of vanishing. Also the sign of the slope of $\Sigma_{2}^{\alpha}(\omega)$ is opposite to what one 
would expect in a conventional Fermi liquid state. More interestingly, corresponding spectral function for orbital 5 and 6 shows no pseudo-gap like behaviour as in 
the case without any SOC effect [see inset of Fig.~\ref{Fig:AlocJH0.01} (c)]. These puzzling features points towards the existence of non-Fermi liquid band.           
\subsection{ARPES}         
\begin{figure}[htbp]
\centering
\includegraphics[scale=0.35,clip=]{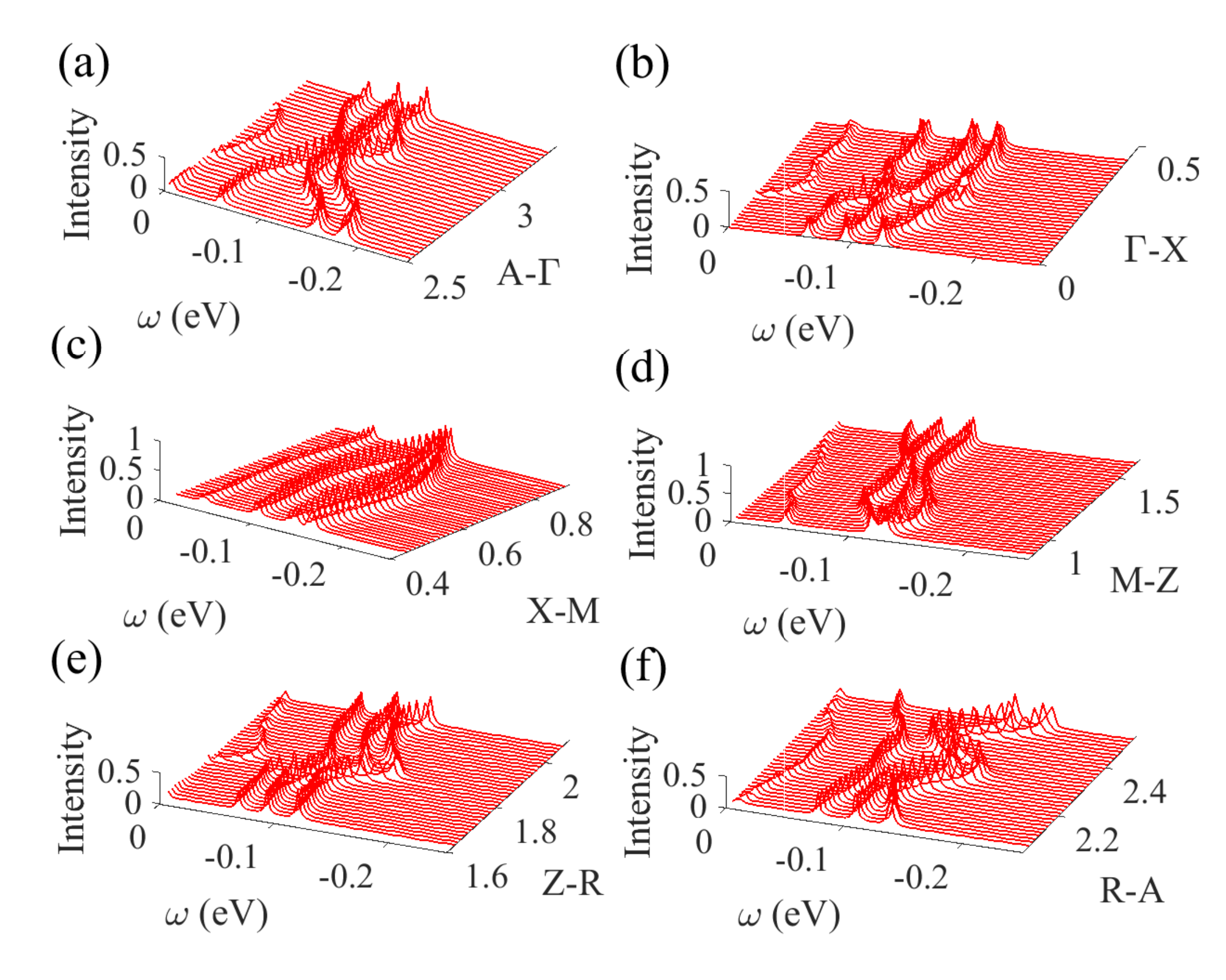}
\caption{(Color online) Spin-orbit coupling effect on ARPES intensity [in arbitary units] spectra at high temperature, $T=0.05\textrm{ eV}$. The spectra is 
calculated along the high symmetry directions : (a) A - $\Gamma$ (b) $\Gamma$ - X (c) X - M (d) M - Z (e)Z - R (f) R - A. The parameters chosen are same as in 
Fig.~\ref{Fig:AlocSOT0.05}.}
\label{Fig:ARPES_SO_T0.05}
\end{figure}
\begin{figure}[htbp]
\centering
\includegraphics[scale=0.35,clip=]{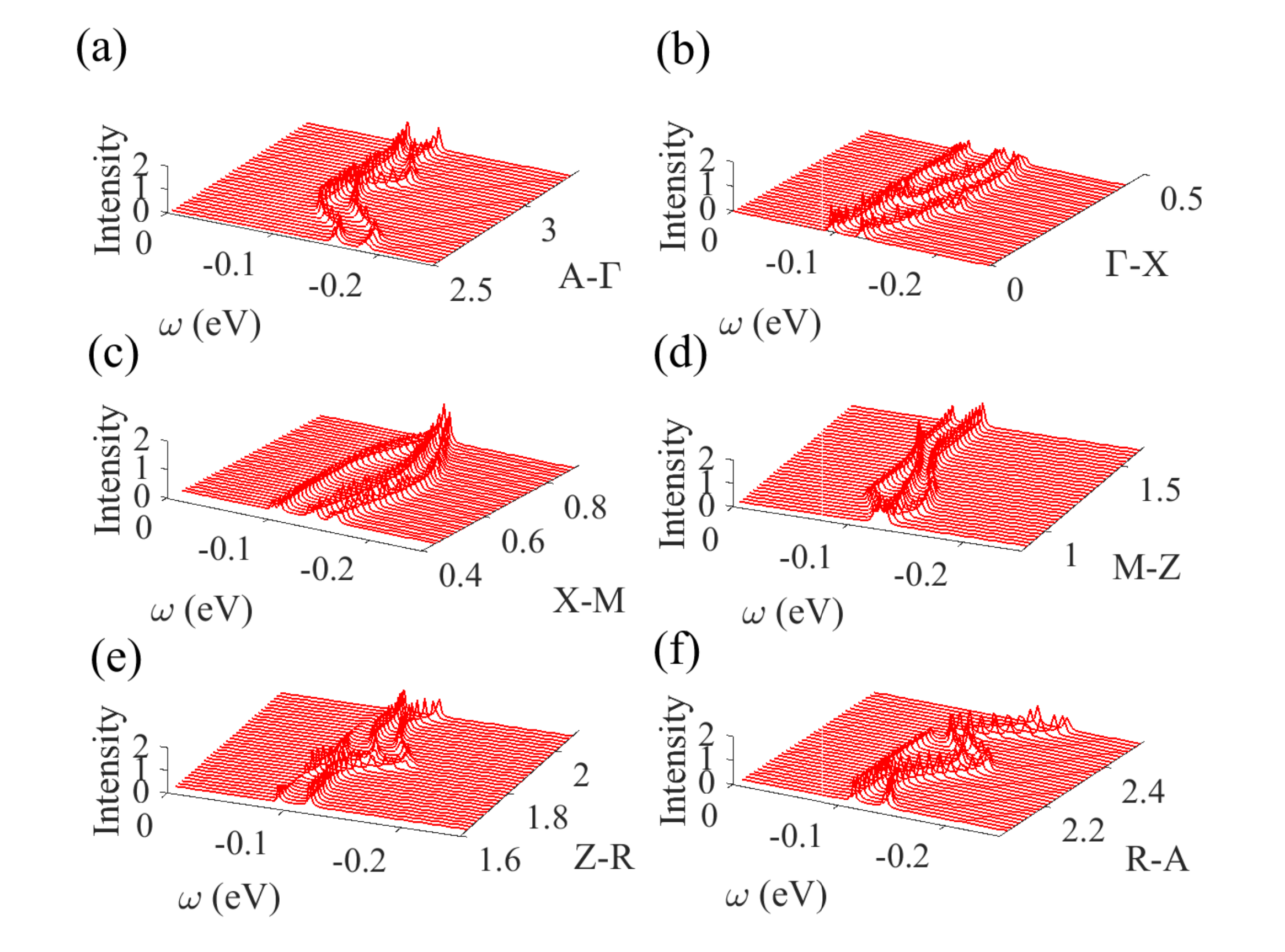}
\caption{(Color online) Spin-orbit interaction effect on ARPES intensity [in arbitary units] spectra at low temperature $T=0.001\textrm{ eV}$. The spectra is 
calculated along the high symmetry directions : (a) A - $\Gamma$ (b) $\Gamma$ - X (c) X - M (d) M - Z (e)Z - R (f) R - A. The parameters chosen are same as in 
Fig.~\ref{Fig:AlocSOT0.05}.}
\label{Fig:ARPES_SO_T0.001}
\end{figure}
In Fig.~\ref{Fig:ARPES_SO_T0.05} we show the effect of SOC on ARPES intensity plot at high temperature $T=0.05\textrm{ eV}$. The first noticeable feature in 
Fig.~\ref{Fig:ARPES_SO_T0.05} is that the spectral features in almost all the symmetry directions are highly dispersive as compared to ARPES intensity plots 
without SOC effect [see Fig~\ref{Fig:ARPES_T0.05}]. This is due to the fact that band dispersion becomes highly dispersive due to the presence of SOC. The spectral 
features arises mainly due to degenerate orbital manifolds (1,2) and (3,4). The contribution of orbital (7,8) is negligible due to Fermi function. Fermi function 
chop off effect is less severe for degenerate manifold (5,6) at high temperature $T=0.05\textrm{ eV}$. The peak just above Fermi energy will appear as a peak 
around $\omega\sim -0.1\textrm{ eV}$. Another noticeable feature is that the convergence of three peak structure into a single peak at $M$ point. This is due to the 
fact that between $X$ and $M$ 4 fold degenerate bands organize into two manifolds and most interestingly the gap between the two manifolds is minimum at $M$-point.

Finally in Fig.~\ref{Fig:ARPES_SO_T0.001} we show the effect of SOC on ARPES at low temperature $T=1\textrm{ meV}$. The spectral features becomes sharp due to 
reduced thermal broadening effects. Also some of the spectral peaks disappears. This is mainly due to Fermi function chopping effect at low temperature as well as 
shifting of self-consistent local chemical potential, $\mu(T)$, to lower energy.     
\subsection{Constant energy surface}
\begin{figure*}
\centering
\includegraphics[width=12cm,clip=]{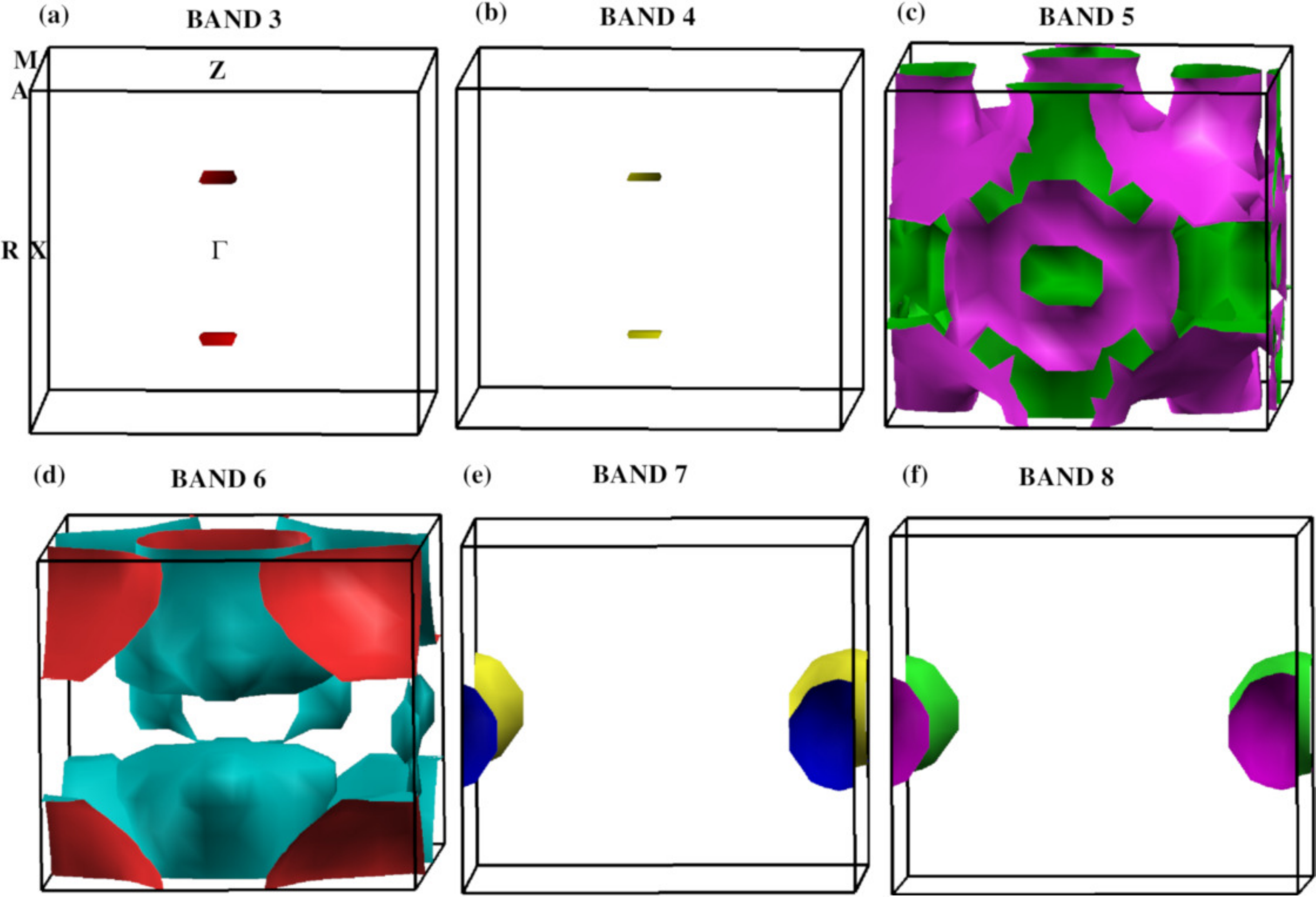}
\caption{(Color online) Effect of spin-orbit coupling on the interacting Fermi surfaces for various orbitals in the first Brillouin zone. The calculation is done 
at $T=1\textrm{ meV}$ and all other parameters are same as in Fig.~\ref{Fig:AlocSOT0.05}}
\label{Fig:FS_int_SO}
\end{figure*}
Finally, we consider strong correlation effects on constant energy surfaces of spin-orbit split bands at low temperature. In Fig.~\ref{Fig:FS_int_SO} we show 
interacting constant zero energy surface of various spin-orbit split bands at $T=1\textrm{ meV}$. As in the case without SOC effect we first solve 
Eq.~\ref{eqn:Ezero} at $T=1\textrm{ meV}$ and then find the constant energy surface $E^{SO}_{\mathbf{k}\alpha}=\varepsilon_{\alpha}^{SO}$ in the first Brillouin 
zone. It is important to mention that in the Fig.~\ref{Fig:FS_int_SO} we do not show interacting zero energy surface for band 1 and 2 (orbital 1 and 2). As these 
orbitals are completely filled and does not cross the Fermi energy. The interacting zero energy surface for band 3 and 4 (orbital 3 and 4) shows dramatic change in 
comparison with non-interacting Fermi surfaces of the corresponding nearly degenerate bands (orbitals). The zero energy surfaces are extremely small and dispersion 
less which will correspond to nearly localized bands (orbitals). These strong correlation effects are also evident in the pseudo-gap like behaviour in the 
corresponding local spectral function as well. The topology of the interacting zero energy surfaces for band 5 and 6 (orbital 5 and 6) shows dramatic change in 
comparison with the corresponding non interating Fermi surfaces. The zero energy surface for band 5 is multiply connected while the zero energy surface of band 6 
consists of simply connected electron and hole pockets. The interacting Fermi surface of band 7 and 8 (orbital 7 and 8) shows no significant change in comparison 
with corresponding nearly degenerate non interacting Fermi surfaces as shown in Fig.~\ref{fig: 5}. 
\section{Effective Model for $\textrm{TmB}_{4}$}  
In the preceding sections we have shown frequency and temperature dependence of spectral function and self-energy for $\textrm{TmB}_{4}$ both in the presence and 
absence of spin-orbit coupling. The imaginary part of self-energy for certain orbitals clearly shows non-Fermi liquid (nFL) characteristics though the corresponding 
spectral functions are gapless at the Fermi energy. In Fig.~\ref{Fig:FS_int} and Fig.~\ref{Fig:FS_int_SO} we show constant zero energy surface of various orbitals in 
the absence and presence of spin-orbit coupling. As we already discussed in the last section that the interacting zero energy surface for certain orbitals 
[see Fig.~\ref{Fig:FS_int_SO}(a) and (b)] show dispersionless and nearly vanishing characteristics in the first Brillouin zone which indicates nearly localized 
nature of the orbitals. Also, we observed that at low temperature the two fold degeneracy between orbital pairs (5,6) is being lifted. The nFL characteristics of 
band 5 and 6 (orbital 5 and 6) is possibly due to unquenched scattering from the nearly localized two fold degenerate orbital pairs (3,4). Interestingly, nearly 
degenerate orbital pairs (7,8) do not show any significant correlation effects [see Fig.~\ref{Fig:FS_int_SO}(e) and (f)] as evident from the nearly unchanged zero 
energy surface in comparison with the corresponding non interacting Fermi surfaces. These two orbitals also show Fermi liquid characteristics [see Fig.~\ref{Fig:ImSigma_SO}(e) and (f)] in their self-energies. The reduced correlation effect is possibly due to much smaller spectral weight of these bands at the Fermi level and hence much weaker scattering from the nearly localized bands (orbitals). Based on these observations, we propose an effective low energy Hamiltonian which will include 
doubly degenerate localized orbitals [corresponding to orbital pair (3,4)], doubly degenerate free band like states [corresponding to orbital pair(7,8)] and doubly 
degenerate orbitals [correspondin to orbital pairs(5,6)] which interacts with the localized orbitals. We propose an effective low energy Hamiltonian for 
$\textrm{TmB}_{4}$ as :
\begin{eqnarray}
&&H_{\textrm{FKM}}= -\sum_{<ij>,\alpha}\left(t^{d}_{\alpha}d^{\dagger}_{i\alpha}d_{j\alpha}+H.c.\right)-\mu\sum_{i,\alpha}\hat{n}_{di\alpha}\nonumber \\
&&\hspace*{2cm}+\sum_{i,\alpha}\left(\epsilon_{\alpha}-\mu\right)\hat{n}_{li\alpha}
+\sum_{i\alpha}U\hat{n}_{li\alpha}\hat{n}_{di\alpha},
\end{eqnarray}     
where $\alpha=1,2$ and $\hat{n}_{li\alpha}\equiv l^{\dagger}_{i\alpha}l_{i\alpha}$, $\hat{n}_{di\alpha}\equiv d^{\dagger}_{i\alpha}d_{i\alpha}$. The model consists 
of two different type of orbital states. At a given lattice site $i$, $l^{\dagger}_{i\alpha}$ ($l_{i\alpha}$) creates (destroys) localized electrons in orbital 
$\alpha$ and onsite energy $\epsilon_{\alpha}$ where as $d^{\dagger}_{i\alpha}$ ($d_{i\alpha\sigma}$) creates(destroys) extended states in orbital $\alpha$ with 
nearest neighbour hopping amplitude $t^{d}_{\alpha}$ and interacts with the localized states through onsite Coulomb correlation $U$. $\mu$ is the chemical potential 
of the system and $\hat{n}_{di\alpha}$, $\hat{n}_{li\alpha}$ are the occupation number operators for $d$ and $l$ states at a given site $i$, respectively. The model 
as described above is essentially a two orbital Falicov-Kimball model. It is important to mention that we do not include free band like states in our Hamiltonian as 
they are merely spectators and their only role is to conserve filling fraction. Also, we do not include spin in our model as in the presence of spin-orbit coupling 
spin is no longer a good qunatum number hence the model described above is a spin less multi orbital Falicov-Kimball model. It is important to mention that a similar 
kind of multi-orbital Falicov-Kimball model was introduced by Taraphder et. al.~\cite{TaraphderPRL2008} as a possible low energy effective model for two dimensional 
colossal magneto-resistive system, $\textrm{GdI}_{2}$. 

It is well known that the metallic state of Falicov-Kimball model is non-Fermi liquid like. The non-Fermi liquid state arises due to unquenched scattering 
from the static species. Similarly, scattering of two band like states from two static species can give rise to two non-Fermi liquid bands. This feature is 
consistent with the spectral properties of $\textrm{TmB}_{4}$. However the Fermi surface for the localized species will completely disappear in contrast to 
observed small [and dispersionless] Fermi surface. This feature can easily be incorporated if we introduce a hopping term for the localized species with hopping 
amplitude $t_{\alpha}^{l}\ll t_{\alpha}^{d}$. However inclusion of such a term will destroy the symmetry of the Falicov-Kimball model and can be neglected under 
lowest order approximation.           
\section{Summary and Conclusion}
To summarize, we have calculated electronic structure of metallic tetra-boride system $\textrm{TmB}_{4}$ using LDA+DMFT scheme. We have studied strong correlation 
effect in this narrow band system by constructing a model Hamiltonian of 4 spin degenerated orbitals which were obtained from DFT calculation [GGA scheme]. From the 
DMFT self-consistency we have calculated spectral function and ARPES along various high symmetry directions. The spectral function consists of a negative energy 
charge transfer peak along with broad peaks around Fermi surface. The integrated spectral weight under one of the orbitals is predominantly above the Fermi energy. 
Hence we can neglect this orbital for thermodynamics as well as low temperature dc transport properties. The imaginary part of self-energy for one of the orbitals 
show strong non-Fermi liquid like characteristics which is possibly linked to pseudo-gap like behaviour of spectral function. It is interesting to mention that 
in multi-orbital systems interplay between strong correlation and Hund's coupling can lead to orbital selective Mott transition(OSMT). The non-Fermi liquid like 
behaviour is possibly a precursor to OSMT. We have also included strong spin-orbit coupling effects present in this system. The inclusion of SOC dramatically 
changes the topological properties of Fermi surface of some of the orbitals. Interestingly degeneracy of orbitals are still preserved which is linked to the nature 
of the magnetic ground state. Inclusion of Hund's coupling effect can lead to degeneracy lifting effects of some degenerate manifold while the other orbitals still 
remains degenerate. The non-Fermi liquid behaviour of certain orbitals are still preserved. Finally we have introduced two orbital Falicov-Kimball model with two 
free band like states as an effective low energy model for $\textrm{TmB}_{4}$. We have discussed in detail the connection of this model with the observed spectral 
properties of the more general 8 band model for $\textrm{TmB}_{4}$.

The interplay of strong spin orbit coupling effect and strong correlation effect in the narrow band system like $\textrm{TmB}_{4}$ will have experimentally 
observable consequences. It is quite natural to expect that in the presence of external magnetic field this system will show Lifshtiz transitions. As a result 
magneto transport properties of this system like longitudinal magneto resistance and Hall effect can show non linear behaviour. The chemical potential tuned by the 
external magnetic filed can lead to partially filled electron like and hole like bands. As a result both semi classical electron like and hole like orbits can 
contribute towards magneto transport. A detailed study on magneto transport of this material is still under investigation.
   
\section*{Acknowledgements}
AT acknowledges Pinaki Sengupta for introducing the material to the group. N.P. would like to acknowledge Sudipta Koley, N. S. Vidyadhiraja, N. Dasari Rao, Monodeep 
Chakraborty, Subhasree Pradhan and Swastika Chatterjee for many stimulating discussions. One of us N. P. would like to acknowledge financial and computational 
support from IIT, Kharagpur.

\end{document}